\documentclass[11pt]{article}
\usepackage[body={16.7cm,23cm}]{geometry}
\usepackage[hang,small,center]{caption}
\usepackage{amsmath,amssymb}
\usepackage{amsfonts}
\usepackage{mathrsfs}
\usepackage{epic}
\usepackage{eepic}
\usepackage{graphicx}
\usepackage{cite}
\usepackage{enumerate}
\usepackage[curve]{xy}
\newcommand{\ds}{\displaystyle}
\renewcommand{\author}[1]{\large\rm #1\\ \bigskip}
\newcommand{\address}[1]{{\normalsize\it #1\\}\bigskip}
\renewcommand{\title}[1]{\bigskip\bigskip\Large\bf #1\bigskip\bigskip\\}

\newcommand{\Bigpsi}[3]{\phantom{\Psi}_2 \kern -.05em
\Psi_2\left(\genfrac{}{}{0pt}{}{#1}{#2}\biggl|#3\right)}

\newcommand{\bea}{\begin{eqnarray}}
\newcommand{\eea}{\end{eqnarray}}

\newcommand{\beq}{\begin{equation}}
\newcommand{\eeq}{\end{equation}}

\newcommand{\ii}{\mathsf{i}}
\newcommand{\oW}{\overline{W}}

\newcommand{\w}{{\mathcal W}}
\newcommand{\s}{{\mathcal S}}

\newcommand{\cpar}{{\eta}}

\newcommand{\lam}{{\lambda}}
\newcommand{\hf}{\frac{1}{2}}
\newcommand{\q}{{\mathsf q}}
\newcommand{\p}{{\mathsf p}}

\newcommand{\G}{{\mathscr G}}
\newcommand{\Lscr}{{\mathscr L}}

\def\EXP{\textrm{{\large e}}}

\def\re{\mathop{\hbox{\rm Re}}\nolimits}
\def\im{\mathop{\hbox{\rm Im}}\nolimits}

\def\Lcal{{\mathcal L}}

\newcommand{\spep}{{p}}
\newcommand{\speq}{{q}}
\newcommand{\sper}{{r}}

\newcommand{\taunew}{\tau'}
\newcommand{\taunewh}{\frac{\tau'}{N}}
\newcommand{\etanew}{\pi}
\newcommand{\xinew}{\phi}
\newcommand{\alphanew}{\theta}

\newcommand{\KM}{{\textsc{(km)}}}

\newcommand{\CP}{{\textsc{(cp)}}}

\newcounter{app}
\newcounter{sapp}[app]
\def\theapp{\Alph{app}}
\newcommand{\app}[1]{
\refstepcounter{app}{\vspace{7mm}
\noindent\Large\bf Appendix
\theapp.
 \ #1 \par \vspace{5mm}}
\setcounter{equation}{0}
\def\theequation{\Alph{app}.\arabic{equation}}}

\renewcommand{\theequation}{{\thesection}.{\arabic{equation}}}
\def\nsection#1{\setcounter{equation}{0}\section{#1}}

\begin{document}

\vglue 2cm

\begin{center}

\title{A master solution of the quantum Yang-Baxter equation\\  and
classical discrete integrable equations}

\author {Vladimir V. Bazhanov}
\address{Department of Theoretical Physics,\\
         Research School of Physics and Engineering,\\
    Australian National University, Canberra, ACT 0200, Australia.\\
E-mail: Vladimir.Bazhanov@anu.edu.au}
\author{Sergey M. Sergeev}
\address{Faculty of Information Sciences and Engineering,\\
University of Canberra, Bruce ACT 2601, Australia.\\
E-mail: Sergey.Sergeev@canberra.edu.au}

\end{center}

\begin{abstract}

We obtain a new solution of the star-triangle relation with positive
Boltzmann weights which contains as special cases
all continuous and discrete spin solutions of this relation,
that were previously
known. This new master solution defines an
exactly solvable 2D lattice model of statistical mechanics,
which involves continuous spin variables,
living on a circle, and contains two temperature-like parameters. If
one of the these parameters approaches a root of unity (corresponds to
zero temperature), the spin
variables freezes into discrete positions, equidistantly spaced on the
circle. An absolute orientation of these positions on the circle
slowly changes between lattice sites by overall rotations. Allowed
configurations of these rotations are described by
classical discrete integrable equations, closely related to the famous
$Q_4$-equations by Adler Bobenko and Suris. Fluctuations
between degenerate ground states in the vicinity of zero temperature
are described by a rather general
integrable lattice model with discrete spin variables.
In some simple special cases the latter reduces to the
Kashiwara-Miwa and chiral Potts models.

\end{abstract}

%\end{document}

\newpage

%\vglue 4.8cm
\nsection{Introduction}
There are only a few exactly solvable models in statistical mechanincs
where the Yang-Baxter equation takes its distinguished ``star-triangular''
form. The most notable discrete-spin models
in this class are the Kashwara-Miwa \cite{Kashiwara:1986}
and chiral Potts \cite{vG85,AuY87,BPA88} models (both of
them also contain the Ising model \cite{O44} and Fateev-Zamolodchikov
$Z_N$-model \cite{FZ82} as particular cases) see \cite{Bax02rip} for
a review.  There are also
important continuous spin models, including Zamolodchikov's
``fishing-net'' model \cite{Zam-fish}, which describes certain planar
Feynman diagrams in quantum field theory, and the Faddeev-Volkov model
\cite{FV95}, connected with quantization \cite{BMS07a} of discrete
conformal transformations \cite{BSp, Steph:2005}. All above models are
also distiguished by the {\em positivity}\/ of the Boltzmann weights ---
the property that is required for many applications, but rarely
realized for generic solutions of the Yang-Baxter equation.

In this paper we present new solutions of the star-triangle relation
which possess the positivity property. Most importantly we present
a {\em master solution}, which contains as special cases
all the previously known continuous and discrete
spin solutions mentioned above\footnote{To be more precise, it only contains
  the solutions, which have a single one-dimensional spin at each lattice
  site. For this reason, it cannot contain the $D\ge2$
fishing-net model which has multi-dimensional spins.},
and also leads to new ones. Our master solution
involves continouos real-valued spins, varying in the range
\ $0\le x,y<2 \pi$, \
and a {\em reflection-symmetric} Boltzmann weight, which is unchanged upon
interchanging the spins $x,y$,
\begin{equation}\label{w-main}
\w_\alpha(x,y)=\w_{\alpha}(y,x)\;=\;%\frac{1}{\kappa(\alpha)}\,
\kappa_\alpha^{-1}\;
\frac{\Phi(x-y+\ii\alpha)}{\Phi(x-y-\ii\alpha)}\;
\frac{\Phi(x+y+\ii\alpha)}{\Phi(x+y-\ii\alpha)}\;,
\eeq
and depends on an {\em additive} spectral parameter $\alpha$ (it
enters additively into the Yang-Baxter equation \eqref{str-main}
below). 
Here $\kappa_\alpha$ is a spin-independent normalization factor and
$\Phi(s)$ is the elliptic $\Gamma$-function
\cite{Ruijsenaars-elliptic,Felder-Varchenko,
Spiridonov-beta}\footnote{%
Our definition \eqref{Phi-def} differs slightly from the standard
definition of the elliptic $\Gamma$-function, see \eqref{our-Phi}
below. The sum in \eqref{Phi-def} is taken over all positive and
negative integers, excluding zero; it converges in the strip\
$|\im{s}\, |<\cpar$, while the product formula \eqref{Gamma}
applies to the whole complex plane of the argument $s$.}\,,
\begin{equation}\label{Phi-def}
\Phi(s)\;=\;\exp\left\{\sum_{n\neq
0} \frac{\EXP^{-\ii
n s}}{n\,(\p^n-\p^{-n})\,(\q^n-\q^{-n})}\right\}\;.
\end{equation}
The latter depends on two fixed parameters $\p$ and $\q$ (elliptic nomes),
\begin{equation}
\p\;=\;\EXP^{\ii\pi\tau}\;,\quad \q=\EXP^{\ii\pi\sigma}\;,\quad
\cpar\;=\;
-\ii\pi\,(\tau+\sigma)\;,\quad \im\tau>0,\quad\im\sigma>0\
.\label{nomes}
\end{equation}
This function obeys a simple functional equation $\Phi(s)\,
\Phi(-s)\equiv1$, which ensures the reflection 
symmetry of the weight \eqref{w-main}.
Define also a single-spin weight
\begin{equation}\label{s-main}
\ds \s(x) \;=\; \frac{\EXP^{{\eta}/{4}}}{4\pi}
\,\vartheta_1(x\,|\,\tau)\,\vartheta_1(x\,|\,\sigma)\;,
\end{equation}
where $\vartheta_j(z\,|\,\tau)$,\  $j=1,2,3,4$,\  are the standard
Jacobi theta-functions \cite{WW} with the periods $\pi$ and $\pi\tau$.

We state that the above weights satisfy the star-triangle relation of
the form,
\beq\label{str-main}
\begin{array}{l}
\ds \int_{0}^{2\pi}\  dx_0 \ \s(x_0)\ {\w}_{\cpar-\alpha_1}(x_1,x_0)\
{\w}_{\alpha_1+\alpha_3}(x_2,x_0)\
\w_{\cpar-\alpha_3}(x_3,x_0)\\[.4cm]
\phantom{\ds\sum_{\sigma} S(\sigma) {W}(\cpar-\theta_1\,|\,a,\sigma)
\,{W}(\cpar-\cpar\cpar)}
={\cal R} \ \w_{\alpha_1}(x_2,x_3)\
\w_{\cpar-\alpha_1-\alpha_3}(x_1,x_3)  \
\w_{\alpha_3}(x_1,x_2),
\end{array}
\eeq
%% \beq\begin{array}{l}
%% \ds \int_{0}^{\pi}\, dx_0 \,\s(x_0)\ {\w}(\cpar-\theta_1\,|\,x_1,x_0)\,
%% \,{\w}(\cpar-\theta_2\,|\,x_2,x_0)\,
%% \w(\cpar-\theta_3\,|\,x_3,x_0)\\[.4cm]
%% \phantom{\ds\sum_{\sigma} S(\sigma) {W}(\cpar-\theta_1\,|\,a,\sigma)
%% \,{W}(\cpar)}
%% ={\cal R}_{123} \ w(\theta_1\,|\,x_2,x_3)\
%% \w(\theta_2\,|\,x_1,x_3)  \
%% \w(\theta_3\,|\,x_1,x_2),
%% \end{array}
%% \label{str-main}
%% \eeq
%% where
%% \beq
%% \theta_1+\theta_2+\theta_3=\cpar,
%% \eeq
%% %
%% and $ {\cal R}_{123}$ is some scalar factor independent of the spins
%% $a, b, c$.
%% %
%% \beq
%% \r_{123}=\Phi(i\cpar-i\theta_1)\  \Phi(i\cpar-i\theta_2)\
%% \Phi(i\cpar-i\theta_3)
%% \eeq
where the {\em crossing parameter} $\cpar$ is defined in
\eqref{nomes} and ${\cal R}$ is some explicitly known scalar
factor (see \eqref{R-definition} below) which depends on the
spectral variables $\alpha_1$ and $\alpha_3$, but is independent of
the spins $x_1,x_2,x_3$.

The weights \eqref{w-main} and \eqref{s-main}
are real
and positive in (at least) two main physical regimes
\beq
\mbox{(i)}\quad \p=\p^*,\quad \q=\q^* \qquad\qquad\mbox{or}\qquad
\mbox{(ii)}\quad \p=\q^* \ ,\label{reg12}
\eeq
with real spins and a real spectral parameter in the range
$0<\alpha<\cpar$. Note, also that the weights are unchanged upon
negating the spins $\w_\alpha(x,y)=\w_\alpha(\pm
x,y)=\w_\alpha(x,\pm y)$, 
\ $\s(x)=\s(-x)$, \ and are
periodic in each spin argument
\beq
\w_\alpha(x,y)=\w_\alpha(x+2\pi,y)=\w_\alpha(x,y+2\pi),\qquad
\s(x)=\s(x+2\pi),\label{w-per}
\eeq
therefore one can regard the spins as angle variables on a circle.
This also means that the integral in \eqref{str-main} 
is a closed contour integral, where the integration contour can be deformed
into the complex plane, if necessary. 

As is well known \cite{Baxterbook} every solution of the
star-triangle relation can be used to define exactly solvable edge
interaction models on various two-dimensional lattices. For purposes of this
introduction it is enough to consider an homogeneous square lattice.
In this case the partition function reads
\beq
{\cal Z}=\mathop{\idotsint}_{0\le x_m<2\pi\phantom{|}}\
\prod_{(ij)}\w_{\alpha}(x_i,x_j)\
\prod_{(kl)}\w_{\cpar-\alpha}(x_k,x_l)\ \prod_{m}
\s(x_m)\,d x_m
\label{z-main}
\end{equation}
where the first product is taken over all horizontal edges $(ij)$, the second
over all vertical edges $(kl)$ and the third over all internal sites
$m$ of the
lattice.
We will implicitly assume fixed boundary conditions.
In the limit of a large lattice the partition function can be
calculated with the inversion relation method \cite{Str79, Zam79,
  Bax82inv}; the result is given in
\eqref{kappa}.

The elliptic $\Gamma$-function \eqref{Phi-def}
first appeared implicitly in Baxter's
pioneering paper \cite{Bax72} on the 8-vertex (it enters the exact
expression for the partition function) and then was
developed systematically in
\cite{Ruijsenaars-elliptic,Felder-Varchenko,
Spiridonov-beta}. Our key observation is that as a mathematical identity
the star-triangle relation \eqref{str-main} reduces
to the most general form of Spiridonov's celebrated
{elliptic beta integral} \cite{Spiridonov-beta}. Actually, it is quite
remarkable that this fundamental integral identity, which lies at
the basis  of the theory of elliptic hypergeometric functions
\cite{Spiridonov-essays}, is nothing but a Yang-Baxter (star-triangle)
relation, defining a perfectly physical integrable lattice model of statistical
mechanics.

Let us now explain how this continuous-spin model could turn into a
model with discrete spin variables, for instance, into the chiral Potts
model. The key to the answer is the {\em low-temperature}\/ limit.
Note that the weights \eqref{w-main} and \eqref{s-main}
symmetrically depend on two temperature-like
parameter $\p$ and $\q$. For generic values
of these parameters, \ $0<|\p|,|\q|<1$, \  the formula
\eqref{w-main} defines  a smooth function of
spins with two dull bell-shaped maxima near $x=y$ and
$x=2\pi-y$. However, when one of the parameters $p,q$ approaches the unit
circle, the maxima become very sharp. Moreover,
the function \eqref{w-main} starts to exhibit additional sharp
($\delta$-function type) maxima, so that the spin variables become
locked to a discrete set of energetically favourable
positions\footnote{Cf. a similar phenomenon for $\q\to-1$ limit in
  ref. \cite{BMS07b}.}. To
get a better idea of how this happens, consider the limit when $\q$
approaches a root of unity
\begin{equation}
\p=\EXP^{i\pi\tau}=\mbox{fixed}, \qquad
\q=\EXP^{-\rho\, T} \EXP^{\ii\pi/N}\;,\qquad T\to
0,\qquad N\ge 1\ , \label{limit}
\end{equation}
and $\rho$ is some suitable chosen numerical coefficient\footnote{%
Note that this limit  does not belong to either of the main physical regimes
\eqref{reg12}, however, with a suitable choice of $\tau$ and $\alpha$
it can be mapped to another physical regime for at least two terms of
the of the low-temperature expansion discussed below.}.
The partition function \eqref{z-main} develops a typical
low-temperature asymptotics
\beq
{\cal Z}= \mathop{\idotsint}_{0\le x_m<2\pi\phantom{|}}\ \
\exp\Big(-\frac{{\cal
    E}\big(X\big)}{T}+O(1)+O(T)\Big)\ \prod_m \frac{d x_m}{\sqrt{
    T}}\ ,\qquad T\to 0\ ,\label{z-low}
\eeq
where $X=\{x_1,x_2,\ldots,x_M\}$ and $M$ is the number of internal
(non-boundary) sites of the lattice.
Explicit expressions for the energy functional
${\cal E}\big(X\big)$ are given in Section~\ref{sec-integ} and
Appendix~\ref{appA}.
It is a bounded from below function of the spin variables
and that for $N>1$ it has stronger periodicity properties in each spin variable
\beq
{\cal
  E}\big(x_1,x_2,\ldots,x_m+{\textstyle\frac{2\pi}{N}}\,,\ldots,x_M\big)=
{\cal E}\big(x_1,x_2,\ldots,x_m\,,\ldots,x_M\big), \quad
m=1,\ldots,M\ ,\label{e-per}
\eeq
than could be expected from \eqref{w-per}. To obtain
the leading asymptotics
of the partition function \eqref{z-low} at $T\to0$, one has to minimize
${\cal E}\big(X\big)$,
\beq
\log{\cal  Z}=-\frac{{\mathcal
  E}(X^{(cl)})}{T}+\log {\cal Z}_0 +O(T)\label{z-exp0},
\eeq
where
\beq
\frac{{\partial {\mathcal E}}\big(X\big)}{\partial
  x_m}\Big\vert_{X=X^{(cl)}}
=0,\qquad m=1,2,\ldots,M\ ,\label{e-min}
\eeq
and $X^{(cl)}=\{\xi_1,\xi_2,\ldots,\xi_M\}$ \ denotes an
stationary point of the functional ${\mathcal E}(X)$,
corresponding to the ground state of the system. Remembering that
the spin variables run over the full circle
$0\le x_m<2\pi$, one immediately
concludes that this ground state is $N^M$- fold degenerate. Indeed,
thanks to \eqref{e-per}, the
equilibrium positions can be independently shifted,
\beq
\xi_m\to\xi_m+{2\pi n_m}/{N}, \qquad n_m\in {\mathbb
  Z}_N\equiv\{0,1,\ldots,N-1\}, \label{flips}
\eeq
without affecting the validity of the conditions
\eqref{e-min}.  Thus at zero temperature, $T=0$,\ the system
``freezes'' into one of these degenerate ground state configurations.

The next natural question is to understand what is happening in
the vicinity of the zero temperature, when the above degeneracy is lifted.
A simple analysis shows that the only fluctuations, that are allowed
in the next-to-leading order of the low-temperature expansion,  are
the discrete flips \eqref{flips} between different ground state
configurations, whereas the values of $\xi_m \pmod
{2\pi/N}$ arising from the minimisation of\  ${\cal E}(X)$ remain frozen.
This means that the quantity ${\cal Z}_0$ in constant term  of the
expansion \eqref{z-exp0} can be understood as
the partition function of a certain edge interaction model with discrete spins
variable $\{n_1,n_2,\ldots,n_M\}$, each taking $N$ different values $n_k\in
{\mathbb Z}_N$. Its Boltzmann weights can be found by expanding
\eqref{z-main} in the limit \eqref{limit} and isolating all constant
terms. The
results are given in Section~\ref{disc-sol}. Naturally,
these Boltzmann weights explicitly depend on the original additive spectral
variables $\alpha$ and $\cpar-\alpha$,
and the remaining temperature-like parameter
$\p$ (which is still at our disposal).
Moreover, they also depend on the variables
$\{\xi_m\}$, which solves the minimization equation \eqref{e-min}, and
this is what makes the problem really complicated. 
In particular, the variables
$\{\xi_m\}$ \ could
depend on the original spectral variables
$\alpha$ and $\cpar-\alpha$ in a very complicated way, therefore,
in general, the simple addition law for the spectral parameters in the
star-triangle relation will be lost. Nevertheless, we definitely know
that the emerging discrete spin model must be integrable! Indeed the
original master model is integrable for any value of parameter $\q$,
therefore this integrability should manifests itself in {\em every
  order} of the expansion \eqref{z-exp0}.

It is clear, of course, that to analytically describe the
new discrete spin  model one
needs to better understand the minimisation equations \eqref{e-min},
which according to our expectations must be integrable as well.
The energy functional ${\cal E}(X)$ is a sum of two-spin edge energies
therefore the variational equation \eqref{e-min} for any particular spin $x_m$
will also involve spins on all neighbouring sites. For the square lattice each
site has four (nearest) neighbours, therefore each equation
\eqref{e-min} will contain five variables.
For the homogeneous model \eqref{z-main} on the
square lattice they have the same form for any internal spin
$\xi$,
 \beq \widetilde{\Psi}_3\big(\xi,\xi_r\big)\widetilde{\Psi}_3(\xi,\xi_\ell)=
\widetilde{\Psi}_1\big(\xi,\xi_u\big)\widetilde{\Psi}_1(\xi,\xi_d)\ ,\label{q4-eq0}
 \eeq
where $\xi_u,\xi_d,\xi_\ell,\xi_r$ are the spins immediately
above, below, to the left and to the right of $\xi$. The function
$\widetilde{\Psi}_j(x,y)$ reads
\begin{equation}\label{psi}
\widetilde{\Psi}_j(x,y)\;=
\;\frac{\vartheta_j\big(\frac{N}{2}(x-y+\ii\alpha)\,|\,N\tau\big)_{\phantom{|}}
  \vartheta_j\big(\frac{N}{2}(x+y+\ii\alpha)\,|\,N\tau\big)}
{\vartheta_j\big(\frac{N}{2}(x-y-\ii\alpha)\,|\,N\tau\big)^{\phantom{|}}
\vartheta_j\big(\frac{N}{2}(x+y-\ii\alpha\,)|\,N\tau\big)}\ ,\qquad
j=1,2,3,4.
\end{equation}
Our next important observation is that these non-linear difference
equations are indeed integrable, as expected.
They appeared previously in remarkable
papers by Adler, Bobenko and Suris \cite{ABS,AS04} devoted to the
classification of classical integrable equations on planar quadrilateral
graphs. More specifically, the minimization equation \eqref{e-min}
arising here is closely related with the so-called $Q_4$ equation,
which is the most complicated equation located at the top of the
Adler-Bobenko-Suris  classification.

The organization of the paper is as follows. In Section~2 we briefly
review some basic facts from the theory of integrable lattice models
and present the master solution \eqref{w-main}, \eqref{s-main}.
The low-temperature expansion is considered in Section~3. In Section~4 we
present a new discrete spin solution of the star-triangle relation and
show that in some simple special cases this solution reduces to those of
the Kashiwara-Miwa \cite{Kashiwara:1986}
and chiral Potts \cite{vG85,AuY87,BPA88} models (the latter
were the most complicated solvable discrete spin models hitherto known).
Here we present only final results; the details of calculations will be
published separately.

\section{The star-triangle relation}
\subsection{Edge interaction models}
\begin{figure}[bt]
\centering
\includegraphics[scale=0.5]{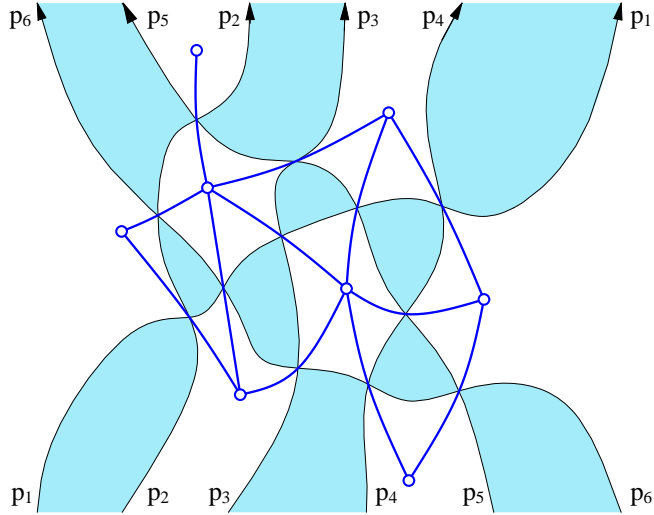}
\caption{The planar graph $\mathscr{G}$ (shown by open circles and bold edges)
and its medial graph $\Lscr$ (shown by thin edges and alternatively
shaded faces). }
\label{fig-net2}
\end{figure}
To facilitate further discussions let us briefly review some basic
facts from the theory of integrable lattice models.
A general solvable edge-interaction model on a
planar graph can be defined in the following way \cite{Bax1,Bax2}.
%.
Consider a planar graph $\G$, of the type shown in Fig.\ref{fig-net2},
where its sites (or vertices) are drawn by open circles and the edges
by bold lines.
The same figure also contains another graph $\Lscr$, shown by thin lines,
which is the {\em medial graph} for $\G$. The faces of ${\Lscr}$ are
shaded alternatively; the sites of $\G$ are placed on the unshaded
faces.
We assume that for each line of $\Lscr$ one can assign a direction,
so that all the lines head generally from
the bottom of the graph to the top.
They can go locally downwards, but there can be no
closed directed paths in $\mathscr{L}$. This means that one can
always distort $\mathscr{L}$, without changing its topology, so
that the lines always head upwards\footnote{%
This assumption puts some restrictions on
the topology of the planar graph $\G$, but still allows enough generality for
our considerations here.}.
%% Let $F({\mathscr G})$, $E({\mathscr G})$
%% and $V({\mathscr G})$ denote the
%% sets of faces, edges and sites of ${\mathscr G}$ and
%% $V_{int}({\mathscr G})$ the set of interior
%% sites of ${\mathscr G}$. The latter correspond to
%% interior faces of $\mathscr{L}$ (i.e., to those having a closed boundary).

Now we define a statistical mechanical model on $\mathscr{G}$.
With each line $\ell$ of ${\mathscr L}$
associate its own ``rapidity'' variable $p_{\ell}$.
At each site $i$ of $\mathscr{G}$ \
place a spin $s_i$, taking some set of (continuous or discrete)
values.
Two spins interact only if they are connected by an edge.  This means
that each edge is assigned with a Boltzmann weight which depends on
spins at the ends of the edge. The Boltzmann weights usually depend on some
global parameters of the model which are same for all edges
(for instance, temperature-like variables).
Here we assume that
they also depend on local parameters, namely, on the
two rapidity variables associated with the edge.
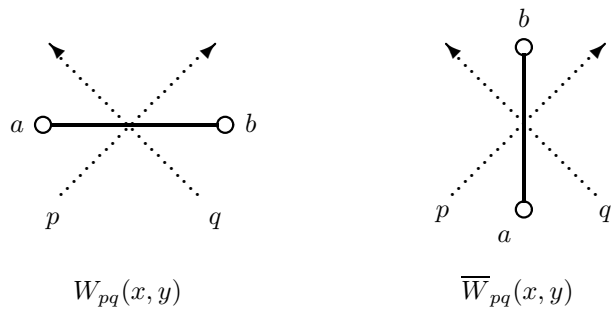
\begin{figure}[hbt]
\begin{center}
\setlength{\unitlength}{.17in} \thicklines
\def\punit#1{\hspace{#1\unitlength}}
\def\pvunit#1{\vspace{#1\unitlength}}
\begin{picture}(4,9)\put(0,1)
{\begin{picture}(4,8)(0,-4.0)
\put(-6.5,0){\line(1,0){5.1}} \put(-3.4,0){\line(1,0){1.0}}
\put(-6.5,-3.2){\makebox(0,0)[b]{\small \mbox{$p$}}}
\put(-1.5,-3.2){\makebox(0,0)[b]{\small \mbox{$q$}}}
\put(-0.4,-0.2){\makebox(0,0)[b]{\small \mbox{$b$}}}
\put(-7.6,-0.2){\makebox(0,0)[b]{\small \mbox{$a$}}}
\put(-4.2,-5.6){\makebox(0,0)[b]{\small \mbox{$W_{pq}(x,y)$}}}
\multiput(-6.4,-2.2)(0.2,0.2){22}{\makebox(0,0)[b]{
\mbox{${.}$}}} \multiput(-6.4,2.0)(0.2,-0.2){22}{\makebox(0,0)[b]{
\mbox{${.}$}}} \put(-6.8,0){\circle{.5}} \put(-1.2,0){\circle{.5}}
\put(8.0,-2.4){\line(0,1){4.6}} \put(8.0,0.3){\line(0,1){1.0}}
\put(5.5,-3.0){\makebox(0,0)[b]{\small \mbox{$p$}}}
\put(10.5,-3.0){\makebox(0,0)[b]{\small \mbox{$q$}}}
\put(8.0,3.0){\makebox(0,0)[b]{\small \mbox{$b$}}}
\put(7.4,-3.6){\makebox(0,0)[b]{\small \mbox{$a$}}}
\put(7.8,-5.6){\makebox(0,0)[b]{\small \mbox{$\oW_{pq}(x,y)$}}}
\multiput(5.8,-2.2)(0.2,0.2){22}{\makebox(0,0)[b]{\mbox{${.}$}}}
\multiput(5.8,2.2)(0.2,-0.2){23}{\makebox(0,0)[b]{\mbox{${.}$}}}
\put(8.0,-2.6){\circle{.5}} \put(8.0,2.4){\circle{.5}}
\put(-1.8,2.2){\vector(1,1){0.3}}
\put(-6.3,2.2){\vector(-1,1){0.3}}
\put(10.2,2.2){\vector(1,1){0.3}}
\put(5.9,2.2){\vector(-1,1){0.3}}
\end{picture}}\end{picture}\end{center}
\caption{Edges of the first (left) and second types and their
  Boltzmann weights.}\label{fig1}
\end{figure}
%The detailed construction is as follows.

\begin{figure}[hbt]
\begin{center}
\includegraphics[scale=0.65]{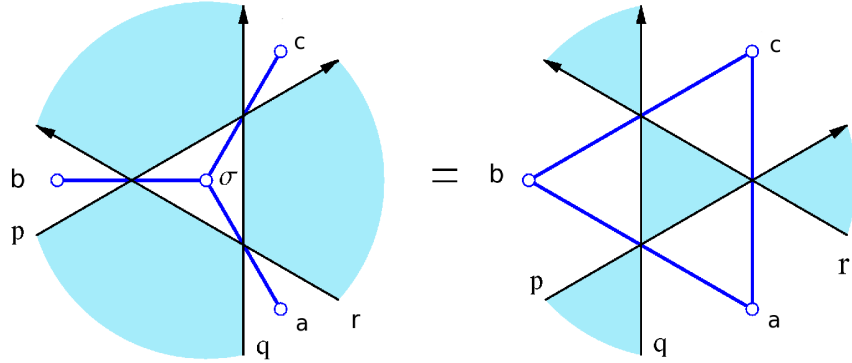}
\end{center}
\caption{A pictorial representation of the first star-triangle relation
 in  \eqref{str-bax}.}
\label{startriangle}
\end{figure}

The edges of $\mathscr{G}$ are either of the first type in
Fig.~\ref{fig1},  or the second. Let $a$, $b$ be the spins at the end
sites of the edge and $p$, $q$ the rapidities of the associated
lines\footnote{%
To avoid confusions note that the rapidity variables $p$ and $q$ are
not related to the elliptic nomes $\p$ and $\q$ in \eqref{nomes},
denoted by upright symbols.},
arranged as in Fig.~\ref{fig1}. Then if the edge is of the first type,
the spins $a$, $b$ interact with the Boltzmann weight function
$W_{pq}(a,b)$. If the edge is of the second type, they interact with
the weight $\oW_{pq}(a,b)$.
In general, there may also be a single-spin self-interaction with a
rapidity-independent weight $S(a)$ for each spin $a$.
The partition function is defined as
\begin{equation}
Z=\sum_{\{s\}}\
\prod_{(ij)}W_{pq}(s_i,s_j)\ \prod_{(kl)}
\oW_{pq}(s_k,s_l) \ \prod_{m} S(s_m)
\label{Z-def}
\end{equation}
where the first product is over all edges $(ij)$ of the first type,
the second is
over all edges $(kl)$ of the second type, and the third one
is over all sites $m$.
The sum is taken over
the values of spins on the internal sites,
while the boundary spins are kept fixed (for continuous spins the sum
is replaced by an integral). The integrability requires the weights to
satisfy the two star-triangle relation \cite{Bax02rip}
\bea
\sum_{d} S(d) \, \overline{W}_{pq}(d,c) W_{pr}(b,d)
\overline{W}_{qr}(a,d) & \! \! =  \! \! &
{R}_{pqr}\,  W_{pq}(b,a)  \overline{W}_{pr}(a,c)
W_{qr}(b,c) \nonumber \\
&& \label{str-bax}\\
\sum_{d} S(d) \, \overline{W}_{pq}(c,d)
W_{pr}(d,b) \overline{W}_{qr}(d,a)  &  \! \! =  \! \!  &
{R}_{pqr}\, W_{pq}(a,b) \overline{W}_{pr}(c,a)
W_{qr}(c,b)  \nonumber
\eea
where  $ { R}_{pqr}$ is some factor independent of the
spins $a, b, c$. For all known solutions
it can be written in the form \cite{MatSmirn90,Bax02rip},
\beq
R_{pqr} =  f_{qr} f_{pq}/f_{pr},\label{Rpqr}
\eeq
where $f_{pq}$ is a scalar function of the two rapidity variables $p$
and $q$. Moreover, the weights can always be normalized so that
they satisfy the two {\em inversion relations},
\bea
\begin{array}{rcl}
\ds\,\sum_c \overline{W}_{pq}(a,c)\,S(c)\,\overline{W}_{qp}(c,b)
&=&S(a)^{-1}\,f_{pq} \,f_{qp}\, \delta_{ab}\ ,
\\[.3cm]
W_{pq}(a,b)\, W_{qp}(a,b)&=&1,
\qquad \forall a,b\ .
\end{array}
\label{inv1}
\eea
The first star-triangle relation in \eqref{str-bax}
equates partition functions of
the ``star'' and ``triangle'' graphs shown in Fig.~\ref{startriangle},
where the external spins $a$, $b$ and $c$ are being fixed. Similarly,
the second relation in \eqref{str-bax} is related to a mirror image of
Fig.~\ref{startriangle}.
Note that the two relations coincide for
reflection-symmetric models,
\beq
W_{pq}(a,b)=W_{pq}(b,a),\qquad
\overline{W}_{pq}(a,b)=\overline{W}_{pq}(b,a),
\label{sym}
\eeq
where the weights are
unchanged by interchanging the spins $a$ and $b$.

The partition function \eqref{Z-def}
possesses remarkable invariance properties \cite{Bax1,Bax2}.
It remains unchanged by continuously deforming the lines of ${\mathscr
  L}$ with their boundary positions kept fixed, as long as the graph
  ${\mathscr L}$ remains directed.
It is easy to see that all such transformations
reduce to a  combination of the moves,
corresponding to the star-triangle \eqref{str-bax} and inversion
relations \eqref{inv1}.
In general the partition function acquire simple  $f_{pq}$
factors under these moves, however with an appropriate
normalization of the Boltzmann weights the invariance is strict (all
extra factors could be eliminated, see \eqref{fr-one}).
Given that the graphs ${\mathscr L}$ and
${\mathscr G}$ can undergo rather drastic
changes, the above invariance, called {\em ``$Z$-invariance''} in
\cite{Bax1},
 is rather non-trivial.
It provides an ultimate formulation of the integrability statement.
For instance, the commutativity of transfer matrices in solvable
models is a particular case of this invariance.

\subsection{The master solution of the star-triangle relation}
The star-triangle relation \eqref{str-main} is a particular case of
\eqref{str-bax}, where
the Boltzmann weights are reflection-symmetric and possess the {\em difference
property}, which means that
they depend only a difference of two rapidities $p$
and $q$,
\beq
W_{pq}(a,b)\to {\cal W}_{p-q}(a,b),\qquad
\overline{W}_{pq}(a,b)\to {\cal W}_{\cpar-p+q}(a,b).\label{difvar}
\eeq
With this correspondence the spectral variables $\alpha_1$ and $\alpha_3$
in \eqref{str-main} are connected to the rapidity variables in
\eqref{str-bax} as
\beq
\alpha_1=q-r,\qquad \alpha_3=p-q\ .
\eeq
Note also that the
fact that the weights for the edges of two types in
Fig.~\ref{fig1} are obtained from each other by a simple
substitution $\alpha\to\cpar-\alpha$\  \ of the difference variable
$\alpha=p-q$ is called the {\em crossing symmetry}. By this reason the
parameter $\cpar$ in \eqref{difvar} is usually called the ``crossing
parameter''.

The standard elliptic $\Gamma$-function is defined as (see, e.g., Eq.(2.18) of
\cite{Spiridonov-essays}),
\begin{equation}\label{Gamma}
\Gamma(z;\p^2,\q^2)\;\stackrel{\textrm{def}}{=}
%\frac{(\p^2\q^2z^{-1};\p^2,q^2)_\infty}{(z;\p^2,q^2)_\infty}=
\;\prod_{n,m=0}^{\infty}
\frac{1-z^{-1}\p^{2n+2}\,\q^{2m+2}}{1-z\,\p^{2n}\,\q^{2m}}\;,
\end{equation}
where
\begin{equation}
\p\;=\;\EXP^{\ii\pi\tau}\;,\quad \q=\EXP^{\ii\pi\sigma}\;,
\end{equation}
and $0<|\p|,|\q|<1$. Define the crossing parameter $\cpar$ by
\begin{equation}
\EXP^{-\eta}\;=\;\p\q\;,\quad \eta \;=\;
-\ii {\pi}(\tau+\sigma)\;.
\end{equation}
For our purposes it is convenient to use the function
\begin{equation}\label{our-Phi}
\Phi(x)\;\stackrel{\textrm{def}}
{=}\;\Gamma(\EXP^{-\ii(x-\ii\eta)};\p^2,\q^2)\;=\;\exp\left\{\sum_{n\neq
0} \frac{\EXP^{-\ii x
n}}{n(\p^n-\p^{-n})(\q^n-\q^{-n})}\right\}\;,
\end{equation}
which satisfies a simple ``reflection'' equation
$\Phi(x)\Phi(-x)=1$.

Our master solution of the star-triangle equation
\eqref{str-main} (already quoted in the
introduction) reads
 \bea\label{ellipt-weight}
\w_{\alpha}(x,y)&=&\kappa(\alpha)^{-1}\,
\frac{\Phi(x-y+\ii\alpha)}{\Phi(x-y-\ii\alpha)}\;
\frac{\Phi(x+y+\ii\alpha)}{\Phi(x+y-\ii\alpha)}\;,\label{w-weight}
\\[.3cm]
\s(x)
%\;=\; \frac{1}{4\pi} \frac{(q^2;q^2)_\infty
%(p^2;p^2)_\infty}{\Gamma(\EXP^{4\ii x})\Gamma(\EXP^{-4\ii
%x})}
&=&\frac{\EXP^{\eta/4}}{4\pi}
\vartheta_1(x\,|\,\tau)\vartheta_1(x\,|\,\sigma)\;.\label{s-weight}
 \eea
where $\vartheta_j(z\,|\,\tau)$,\  $j=1,2,3,4$,\  are the standard
Jacobi theta-functions \cite{WW} with the periods $\pi$ and
$\pi\tau$. The weights are symmetric,
 \beq
\w_\alpha(x,y)=\w_\alpha(y,x)=\w_{\alpha}(\pm x,y)=
\w_{\alpha}(x,\pm y)\,, \qquad
\s(x)=\s(-x),\label{sym1} \eeq periodic in their spin arguments,
\beq \w_\alpha(x,y)=\w_\alpha(x+2\pi,y)=\w_\alpha(x,y+2\pi),\qquad
\s(x)=\s(x+2\pi), \label{w-per2}
 \eeq
and satisfy the recurrence relations
\begin{equation}\label{sl2-difference}
\frac{\w_\alpha\big(x-{\pi\sigma},y\big)_{\phantom{|}}}
{\w_\alpha\big(x+\pi\sigma,y\big)^{\phantom{|}}}\;=\;
\frac{\vartheta_4\big(\hf(x-y+\ii\alpha)\,|\,\tau\big)_{\phantom{|}}}
{\vartheta_4\big(\hf(x-y-\ii\alpha)\,|\,\tau\big)^{\phantom{|}}}
\frac{\vartheta_4\big(\hf(x+y+\ii\alpha)\,|\,\tau\big)_{\phantom{|}}}
{\vartheta_4\big(\hf(x+y-\ii\alpha)\,|\,\tau\big)^{\phantom{|}}}\;,\quad
\textrm{and the same  with}\ \tau\leftrightarrow\sigma\;.
\end{equation}
For real-valued spins the
weights are real and positive in the two main physical regimes
\beq
\mbox{(i)}\quad \p=\p^*,\quad \q=\q^* \qquad\qquad\mbox{or}\qquad
\mbox{(ii)}\quad \p=\q^* \ ,\label{reg12-2}
\eeq
provided the spectral parameter is real and
kept in the range $0<\alpha<\cpar$.

As a mathematical identity the star-triangle relation
\eqref{str-main} contains seven continuous parameters: two nomes
$\p$ and $\q$; two spectral variables $\alpha_1$ and $\alpha_3$; and
three spins $x_1,x_2,x_3$. One can show that this
relation can be reduced to the most general form of Spiridonov's
celebrated elliptic beta integral. Then using Eq.(3.1) of
\cite{Spiridonov-essays}  one obtains the expression for the
factor ${\cal R}$ in \eqref{str-main},
 \beq\label{R-definition} {\cal
R}=\frac{f(\alpha_1)\,f(\alpha_3)}{f({\alpha_1+\alpha_3})},\qquad
{f(\alpha)}=\Phi(\ii\cpar-2\ii\alpha)\,\frac{\kappa(\alpha)}
{\kappa(\cpar-\alpha)}\ ,
 \eeq
It is worth noting that if the
function $\kappa(\alpha)$ satisfies the equations
\begin{equation}
\frac{\kappa(\eta-\alpha)}{\kappa(\alpha)}\;=\;\Phi(\ii\cpar-2\ii\alpha)
\;,\quad
\kappa(\alpha)\kappa(-\alpha)=1\;,\label{invfunc}
\end{equation}
then one trivially obtains
\beq
f(\alpha)\equiv1,\qquad {\cal R}\equiv1\ .\label{fr-one}
\eeq
With this normalization the inversion relations \eqref{inv1} for the weights
\eqref{w-weight} and \eqref{s-weight} simplify to
\bea
\int_0^{2\pi} dz \,\s(z)\,\w_{\eta-\alpha}(x,z)\,\w_{\eta+\alpha}(z,y)&=&\frac{1}{2\s(x)}
\left(\delta(x-y)+\delta(x+y)\right)\;,\nonumber
\\
\label{inv2}
\\
\w_\alpha(x,y)\,\w_{-\alpha}(x,y)&=&1\,,\nonumber
\eea
where $\delta(x)$ denotes the periodic $\delta$-function
\beq
\delta(x)=\frac{1}{2\pi}\sum_{n=-\infty}^\infty \EXP^{\ii n x}\ .
\eeq
Moreover, for the same normalization the weights have the following special 
values
\beq
\w_\alpha(x,y)\Big\vert_{\alpha=0}=1,\qquad 
\w_{\cpar-\alpha}(x,y)\Big\vert_{\alpha\to+0}=\frac{1}{2\s(x)}
\left(\delta(x-y)+\delta(x+y)\right)
\eeq
Note that all the above relations are consistent with
the symmetries of the weights \eqref{sym1}.

The general definition of the partition function \eqref{Z-def} specialized  
to this case leads to
\begin{equation}
{\cal Z}=\int
\prod_{(ij)}{\cal W}_{\alpha_{ij}}(x_i,x_j)\ \prod_{m} {\cal S}(x_m)\ dx_m
\label{Z-def2}
\end{equation}
where the edge variables $\alpha_{ij}$
are given by
\beq
\alpha_{ij}=\left\{
\begin{array}{ll}
p-q,\qquad &\mbox{for a first type edge}\\[.3cm]
\cpar-p+q,\qquad &\mbox{for a second type edge}
\end{array}\right.\label{rule}
\eeq
where $p$ and $q$ are the pair of rapidities,
associated the edge $(ij)$, as shown in Fig.~\ref{fig1}.
 The integral in \eqref{Z-def2} is over all
internal spins. The boundary spins are kept fixed (e.g., all boundary
spins can be set to zero).

The equations \eqref{invfunc} are precisely the functional equation
of the inversion relation method \cite{Str79, Zam79,Bax82inv} adapted for
edge interaction models \cite{BS10b}. The quantity $\kappa(\alpha)$
is interpreted there as the partition function per edge in the limit of an
infinitely large lattice. A suitable solution of \eqref{invfunc} with
appropriate analytic   properties is given by
\begin{equation}
\kappa(\alpha)\;=\;\exp\left\{\sum_{n\neq 0}
\frac{\EXP^{2\alpha
n}}{n(\p^n-\p^{-n})(\q^n-\q^{-n})(\p^n\q^n+\p^{-n}\q^{-n})}\right\}\;.\label{kappa}
\end{equation}
If this function is used for the normalization of the Boltzmann weights
\eqref{ellipt-weight} in \eqref{Z-def2} then the partition function per
site
\beq
\lim_{M\to\infty} {\cal Z}^{1/M}=1,
\eeq
is equal to one in the thermodynamic limit\footnote{Given that the
  inversion relation method \cite{Str79, Zam79,Bax82inv} require
  analyticity assumptions, which cannot be rigorously justified, it
  would be desirable to reproduce the result \eqref{kappa} by other
  methods, e.g., by a
 kind of Bethe Ansatz.}.

 \section{Low-temperature expansion}
\subsection{Asymptotics of the weights}
Consider the limit of star-triangle equation \eqref{str-main}
when one of the temperature-like
parameters tends to a root of unity,
\begin{equation}
\p=\EXP^{i\pi\tau}=\mbox{fixed}, \qquad \q=\EXP^{-\varepsilon}
\EXP^{\ii\pi/N}\;,\qquad \varepsilon\to 0,\qquad N\ge 1\ .
\label{limit2}
\end{equation}
Note that
the crossing parameter \eqref{nomes} then becomes
\beq
\label{eta-lim}
\cpar_\varepsilon=-\ii\,\Big(\frac{\pi}{N}+\pi\tau
+\frac{\ii\varepsilon}{2 N^2}\Big)
 \to
\cpar=-\ii\, \Big(\frac{\pi}{N}+\pi\tau\Big)\ .
\eeq
As explained in the Introduction the Boltzmann weights develop
a singular asymptotics \eqref{z-low} where the leading term
is unchanged upon the shifts $x_{i}\to
x_i+2\pi/N$. Therefore it is convenient to set
\beq
x_i=\xi_i+\frac{2\pi n_i}{N}\ ,\quad n_i\in{\mathbb Z}_N,
\quad -\frac{\pi}{N}<\re \xi_i<\frac{\pi}{N}.\label{positions}
\eeq
From now on we will use new variables
\beq\label{newvars}
\alphanew=
\frac{\ii N \alpha}{1+N\tau},\quad \xinew_i=\frac{N}{1+N\tau}\Big(
\xi_i+\frac{\pi\tau}{2}\Big),
\quad \tau'=\frac{N\tau}{1+N\tau},\quad \eta'=\frac{\ii N
  \cpar}{1+N\tau}=\pi\ .
\eeq
instead of $\alpha$, $\xi_i$, $\tau$ and $\eta$. 
With the new variables the asymptotics of \eqref{w-weight} and
  \eqref{s-weight} can be written as\footnote{See Appendix A for the
    corresponding expressions in the original variables}
\bea
\log {\cal W}_\alpha(x_1,x_2)&=&-\frac{1}{\varepsilon}\,{\cal
    L}(\alphanew\,|\,\phi_1,\phi_2)
+\log W_\alphanew(\xinew_1,\xinew_2,n_1,n_2)
+O(\varepsilon)\ ,\label{w-ass}\\[.3cm]
\log \s(x_i)&=&-\frac{1}{\varepsilon}\,{\cal C}(\xinew_i)
-\frac{1}{2}\log{\varepsilon}+\log S(\xinew_i,n_i) +O(\varepsilon)\label{s-ass}
\eea
where
\begin{equation}\label{Cphi}
{\cal C}(\phi_1)=\frac{1}{2}\,\left(\frac{2\phi_1-\pi}{1-\tau'}\right)^2
\end{equation}
and
\beq
\label{Lag1}
\begin{array}{l}
\ds{\cal
  L}(\theta|\phi_1,\phi_2)=
-\frac{\theta}{2\pi}\Big({\cal C}(\phi_1)+{\cal C}(\phi_2)\Big)
\\[.5cm]
\ \ \ \ \ \ \ds+\frac{\ii N\tau }{\tau'}
\left\{
\mathop{\int}_0^{\ \ \phi_1-\phi_2}dz
\log\frac{\vartheta_2\big(\frac{1}{2}(z-\theta)\,|\,\tau'\big)}
{\vartheta_2\big(\frac{1}{2}(z+\theta)\,|\,\tau'\big)}
+\mathop{\int}_\pi^{\ \ \phi_1+\phi_2}dz
\log\frac{\vartheta_3\big(\frac{1}{2}(z-\theta)\,|\,\tau'\big)}
{\vartheta_3\big(\frac{1}{2}(z+\theta)\,|\,\tau'\big)}
\right\}
\end{array}
\eeq
Note the leading terms in \eqref{w-ass} are independent of the
integers $n_i$, entering \eqref{positions}. 
Explicit expressions
for  $W_\alphanew(\xinew_1,\xinew_2,n_1,n_2)$ and $S(\xinew,n)$ are
given in the next Section.

\subsection{Expansion of the star-triangle relation}
In deriving \eqref{w-ass}
we assumed the normalization \eqref{kappa} for which the factor ${\cal
  R}$ in \eqref{str-main} is equal to one. Substituting \eqref{w-ass}
and \eqref{s-ass} into the star-triangle relation \eqref{str-main},
one obtains
\begin{equation}\label{str-int}
\begin{array}{l}
\ds
\int\frac{d\xinew_0}{\sqrt{\varepsilon}}\,
%\EXP^{\ds-\frac{\mathcal{A}_\star(\xinew)}{\varepsilon}}
\:\EXP^{\ds-{\mathcal{A}_\bigstar(\xinew)}/{\varepsilon}}
\ \Bigg\{ \sum_{{n_0}\in{\mathbb Z}_N}
S(n_0) \,W_{\etanew-\alphanew_1}(n_1,n_0)\,
W_{\alphanew_1+\alphanew_3}(n_2,n_0)\,
W_{\etanew-\alphanew_3}(n_3,n_0)\Bigg\}
\\[1.2cm]
\ds \phantom{xxxxxxxx} =\ds(\tau'/\tau)\,
\EXP^{\ds{-{\mathcal{A}_\vartriangle(\xinew)}/{\varepsilon}}}
\ \
W_{\alphanew_1}(n_2,n_3)\, W_{\pi-\alphanew_1-\alphanew_3}(n_1,n_3)\,
W_{\alphanew_3}(n_1,n_2)\,\big(1+O(\varepsilon)\big)\,,
\end{array}
\end{equation}
where $\xinew=(\xinew_0,\xinew_1,\xinew_2,\xinew_3)$,\  and
\bea
\mathcal{A}_{\bigstar}(\xinew)&=&
\Lcal(\etanew-\theta_1\,|\,\xinew_1,\xinew_0)+
\Lcal(\alphanew_1+\alphanew_3\,|\,\xinew_2,\xinew_0)
+\Lcal(\etanew-\theta_3\,|\,\xinew_3,\xinew_0)+{\cal C}(\xinew_0)\;,\label{A-star}\\[.3cm]
\mathcal{A}_{\triangle}(\xinew)&=&\Lcal(\theta_1\,|\,\xinew_2,\xinew_3)+
\Lcal(\etanew-\alphanew_1-\alphanew_3\,|\,\xinew_3,\xinew_1)
+\Lcal(\theta_3\,|\,\xinew_1,\xinew_2)\;.\label{A-tri}
\eea
Here we used the abbreviated notations $S(\xinew_0,n_0)\equiv S(n_0)$,
\  $W_\alphanew(n_i,n_j)\equiv
W_{\alphanew}(\xinew_i,\xinew_j,n_i,n_j)$, assuming an implicit dependence on
the variables $\xinew_i$.

Evaluating the integral \eqref{str-int} by the saddle point method
one immediately obtains two non-trivial identities
valid for arbitrary values of $\xinew_1,\xinew_2,\xinew_3$.
The first of these relations (in the leading order in
$\varepsilon$) reads
\beq
\mathcal{A}_\bigstar (\xinew_0^{(cl)},\xinew_1,\xinew_2,\xinew_3) =
\mathcal{A}_\triangle(\xinew_1,\xinew_2,\xinew_3), \label{cstr}
\eeq
where $\xinew_0^{(cl)}$ is the stationary point of the integral in
\eqref{str-int}, i.e., the value of $\xinew_0$, which solves the equation
\begin{equation}
\frac{\partial{\mathcal A}_\bigstar(\xinew)}
{\partial \xinew_0}\Big|_{\xinew_0=\xinew_0^{(cl)}}
=0\ .\label{statpoint}
\end{equation}
In the following we will omit the superfix ``(cl)'' and always assume
that $\xinew_0\equiv \xinew_0^{(cl)}$.
To write \eqref{statpoint} explicitly, define the function
\begin{equation}\label{newpsi}
\Psi_\alphanew\big(\xinew_i,\xinew_j\big)=
\frac{\vartheta_2\big(\frac{1}{2}
\big(\xinew_i-\xinew_j+\alphanew\big)\,|\,\taunew\big)}
{\vartheta_2\big(\frac{1}{2}
\big(\xinew_i-\xinew_j-\alphanew\big)\,|\,\taunew\big)}
\frac{\vartheta_3
\big(\frac{1}{2}\big(\xinew_i+\xinew_j+\alphanew\big)\,|\,\taunew\big)}
{\vartheta_3
\big(\frac{1}{2}\big(\xinew_i+\xinew_j-\alphanew\big)\,|\,\taunew\big)}\;.
\end{equation}
Using the expressions for ${\cal L}(\alphanew\,|\,\xinew_1,\xinew_2)$
given \eqref{Lag1} one can write the equation \eqref{statpoint} in the
product form
\begin{equation}\label{Q4}
\Psi_{\pi-\alphanew_1}\big(\xinew_0,\xinew_1\big)\,
\Psi_{\alphanew_1+\alphanew_3}\big(\xinew_0,\xinew_2\big)\,
\Psi_{\pi-\alphanew_3}\big(\xinew_0,\xinew_3\big)\;=\;1\;.
\end{equation}

The second relation, which follows from \eqref{str-int}
(in the order $O(\varepsilon^0)$) reads
\begin{equation}\label{str-sum}
\begin{array}{l}
\ds \sum_{n_0\in\mathbb{Z}_N} S(\phi_0,n_0)
W_{\etanew-\alphanew_1}(\xinew_1,\xinew_0,n_1,n_0)
W_{\alphanew_1+\alphanew_3}(\xinew_2,\xinew_0,n_2,n_0)
W_{\etanew-\alphanew_3}(\xinew_3,\xinew_0,n_3,n_0)\\
[5mm]
\ds \phantom{xxxxxx} \;=\;
R\
W_{\alphanew_1}(\xinew_2,\xinew_3,n_2,n_3)\,
W_{\pi-\alphanew_1-\alphanew_3}(\xinew_1,\xinew_3,n_1,n_3)\,
W_{\alphanew_3}(\xinew_1,\xinew_2,n_1,n_2)\;,
\end{array}
\end{equation}
where $\xinew_0,\xinew_1,\xinew_2,\xinew_3$ satisfy the equations
\eqref{Q4} and
\beq
R=(\tau'/\tau)\,\big({\mathcal{A}''_{\bigstar}(\xinew_0)}/{2\pi}\big)^\hf\,
\label{Rsimple}\ .  \eeq
The quantity ${{\cal A}''_\bigstar(\xinew_0)}$ denote the second derivative of the
action \eqref{A-star} at the stationary point \eqref{statpoint}.
\subsection{Energy functional}
\label{sec-integ}
Using the asymptotics \eqref{w-ass}, \eqref{s-ass} in \eqref{Z-def2}
and calculating the integral by the saddle point method one obtains
\beq
\log {\cal Z}=-\frac{1}{\varepsilon}\,{\cal
  E}(\phi)+O(1)+O(\varepsilon)\,. \label{z-ass3}
\eeq
The energy functional reads
\beq
{\cal E}(\phi)=\sum_{(ij)} {\cal L}(\alphanew_{ij}\,|\,\phi_i,\phi_j)
+\sum_m{\cal C}(\phi_m)\ ,\label{E2}
\eeq
where
\beq
\theta_{ij}={\ii\, \alpha_{ij}N}/({1+N\tau})\,,
\eeq
and the variables $\phi=\{\phi_1,\phi_2,\ldots,\phi_M\}$ solve the
variational equations
\beq
\prod_j\,\Psi_{\theta_{ij}}(\phi_i,\phi_j)=1,\quad i=1,2,\ldots,M\label{vareq}
\eeq
where for any internal site
$i$ the product is taken over all edges ${(ij)}$ meeting
at $i$ (the index $j$ numerates these edges). The function $\Psi$ is
defined in \eqref{newpsi}.  Note a useful sum rule
\cite{BMS07a} which is a corollary of the definition \eqref{rule} for
lattices of the type described in Section~2.1,
\beq
\sum_j\theta_{ij}=2\pi\ .
\eeq
Note that by this property the ${\cal C}(\phi_i)$ terms, 
that arise from the first sum in \eqref{E2} (see \eqref{Lag1}), exactly 
cancel out the second sum in \eqref{E2}.

The equation \eqref{cstr} is the {\em classical star-triangle
relation}.  It was introduced in \cite{BMS07a} and used there to
prove the invariance of energy (or action) functionals of the type
\eqref{E2} under deformations of the lattice, discussed in Section~2.1,
namely under the star-triangular moves (this is the classical analog
of Baxter's $Z$-invariance). Recently the 
equation \eqref{cstr} was discussed in \cite{BS09} in connection with
$Q_4$ equation and other integrable equations from \cite{ABS} (see
also \cite{Lobb09a}). As noted in \cite{BS09} the stationarity equation
\eqref{Q4} could be identified with the so-called {\em three-leg form}
  of the $Q_4$-equations, which could be written in other equivalent forms.

Differentiating \eqref{cstr}
with respect to $\xinew_j$, and taking into account \eqref{statpoint},
\beq
\frac{\partial}{\partial \xinew_j}
\Big(
\mathcal{A}_\bigstar (\xinew_0,\xinew_1,\xinew_2,\xinew_3)-
\mathcal{A}_\triangle(\xinew_1,\xinew_2,\xinew_3)\Big)=0,\qquad
j=1,2,3,\label{trileg}
\eeq
one obtains another three equivalent forms of \eqref{statpoint}.
These equations can be written similarly to
\eqref{Q4},
\begin{equation}\label{Q4-others}
\begin{array}{rcl}
\ds \Psi_{\pi-\alphanew_1}(\xinew_1,\xinew_0)&=&
\Psi_{\pi-\alphanew_1-\alphanew_3}(\xinew_1,\xinew_3)
\Psi_{\alphanew_3}(\xinew_1,\xinew_2)\;,\\
[5mm]
\ds
\Psi_{\alphanew_1+\alphanew_3}(\xinew_2,\xinew_0)&=&
\Psi_{\alphanew_1}(\xinew_2,\xinew_3)
\Psi_{\alphanew_3}(\xinew_2,\xinew_1)\;,\\
[5mm]
\ds \Psi_{\pi-\alphanew_3}(\xinew_3,\xinew_0)&=&
\Psi_{\alphanew_1}(\xinew_3,\xinew_2)
\Psi_{\pi-\alphanew_1-\alphanew_3}(\xinew_3,\xinew_1)\;,
\end{array}
\end{equation}
All these relations can be brought to the canonical form of the
$Q_4$-equation from the Adler-Bobenko-Suris list \cite{ABS} by the
substitution
\begin{equation}
u_j=\frac{\vartheta_1(\frac{1}{2}\xinew_j\,|\,\frac{1}{2}\taunew)}
{\vartheta_2(\frac{1}{2}\xinew_j\,|\,\frac{1}{2}\taunew)}\;,\quad
j=0,1,3\;,\quad
u_2=\frac{\vartheta_2(\frac{1}{2}\xinew_2\,|\,\frac{1}{2}\taunew)}
{\vartheta_1(\frac{1}{2}\xinew_2\,|\,\frac{1}{2}\taunew)}\;,\quad
\textrm{and}\quad
T(\alphanew)\;=\;\frac{\vartheta_1(\frac{1}{2}\alphanew\,|\,\frac{1}{2}\taunew)}
{\vartheta_2(\frac{1}{2}\alphanew\,|\,\frac{1}{2}\taunew)}\;.
\end{equation}
Then, each of Eqs. \eqref{Q4} and (\ref{Q4-others}) can be reduced to
\begin{equation}\label{my-Q4}
\begin{array}{l}
\ds T(\alphanew_1)\,(u_0u_1-u_2u_3) + T(\alphanew_1+\alphanew_3)\,
(u_0u_2-u_1u_3) +
T(\alphanew_3)\,(u_0u_3-u_1u_2)\\
[5mm]
\ds \ \ \ \ \ \ \ \ \ 
+ T(\alphanew_1)\,T(\alphanew_1+\alphanew_3)\,T(\alphanew_3)\,
(u_0u_1u_2u_3-1)=0\;.
\end{array}
\end{equation}
This is an ``affine-linear'' constraint on four variables
$u_0,u_1,u_2,u_3$, which can be solved for each
                                                 of them via
simple rational functions of the other three.

\section{Discrete spin solutions of the star-triangle equation}
\label{disc-sol}
Before proceeding to the details of the solution of the star-triangle
equation \eqref{str-sum}, let us explain some conceptual modification
to the integrability scheme in this case. We need to consider an arbitrary
planar graph of the type discussed in Section~2.1 and assign two types
of spin variables
to the lattice sites: classical variables $\phi_i$, satisfying equations
\eqref{vareq}, and discrete spin variables $n_i\in{\mathbb Z}_N$.
Then one needs to solve all equations \eqref{vareq} and determine all
$\phi_i$. After that the discrete spin model is defined by the weights
$W_\theta(\phi_i,\phi_j,n_i,n_j)$ and $S(\xinew_0,n_0)$ given by
\eqref{ellipt-cw}. Previously, related ideas were developed in 
\cite{BBR,Baz08}.

It is important to note that the Baxter's $Z$-invariance property
manifests itself in each order of the 
lower temperature expansion. Namely, the classical
star-triangle relation \eqref{cstr} ensures that 
classical action \eqref{E2} is invariant under the star-triangle moves
of the lattice. Likewise the ``quantum'' star-triangle relation
\eqref{str-sum} and the variational equations \eqref{vareq} ensure
this invariance for the partition function of the discrete spin
model defined by \eqref{ellipt-cw}.

\subsection{General solution}
Here we present explicit expressions for the weights solving the
star-triangle relation \eqref{str-sum} that arise  in the low-temperature
limit, described in the previous section. The value of the factor $R$
given in \eqref{Rsimple} corresponds to the normalization of weights
inherited from  \eqref{w-weight}, \eqref{kappa} via the expansion
\eqref{w-ass}. This normalization it not particularly natural.
Here we will use the much more conventional normalization
\beq
W_\alphanew(\xinew_i,\xinew_j,0,0)=1,\qquad \forall \xinew_i,\xinew_j
\ .\label{newnorma}
\eeq
The corresponding value of the factor $R$ is given in \eqref{Rnew} below.
Define two functions
\begin{equation}
r_{\alphanew}^{}(\xinew;n)\;=\;\left[\frac{\vartheta_2(\frac{1}{2}(\xinew+\alphanew)\,|\,\taunew)}
{\vartheta_2(\frac{1}{2}(\xinew-\alphanew)\,|\,\taunew)}\right]_{}^{n/N}
\prod_{k=1}^n\frac{\vartheta_1\big(\frac{\pi}{N}(k-\frac{1}{2})+\frac{1}{2N}(\xinew-\alphanew)\,|\,\taunewh\big)}
{\vartheta_1\big(\frac{\pi}{N}(k-\frac{1}{2})+\frac{1}{2N}(\xinew+\alphanew)\,|\,\taunewh\big)}\;,
\end{equation}
and
\begin{equation}
t_{\alphanew}^{}(\xinew;n)\;=\;\left[\frac{\vartheta_3(\frac{1}{2}(\xinew+\alphanew)\,|\,\taunew)}
{\vartheta_3(\frac{1}{2}(\xinew-\alphanew)\,|\,\taunew)}\right]_{}^{n/N}
\prod_{k=1}^n\frac{\vartheta_4\big(\frac{\pi}{N}(k-\frac{1}{2})+\frac{1}{2N}(\xinew-\alphanew)\,|\,\taunewh\big)}
{\vartheta_4\big(\frac{\pi}{N}(k-\frac{1}{2})+\frac{1}{2N}(\xinew+\alphanew)\,|\,\taunewh\big)}\;.
\end{equation}
They possess the following symmetries
\beq
r_\alphanew(\xinew;n+N)=r_{\alphanew}(-\xinew,-n)=r_{\alphanew}(\xinew;n),
\quad
t_\alphanew(\xinew;n+N)=t_{\alphanew}(-\xinew,-n)=t_{\alphanew}(\xinew;n),
\eeq
and $r_\alphanew(\xinew;0)=t_\alphanew(\xinew;0)=1$.
These functions generalize those used in \cite{Bax02rip} in connection
with the Kashiwara-Miwa model.

Then the weights satisfying (\ref{str-sum}) are given
by
\begin{equation}\begin{array}{l}\label{ellipt-cw}
\ds W_{\alphanew}(\xinew_i,\xinew_j;n_i,n_j)\;=\;
r_{\alphanew}(\xinew_i-\xinew_j;n_i-n_j) t_{\alphanew}(\xinew_i+\xinew_j;n_i+n_j)\;,\\
[3mm]
\ds
S(\xinew_0,n_0)\;=\;\frac{1}{\sqrt{N}}
\frac{\vartheta_4(\frac{2\pi}{N}n_0+\frac{1}{N}\xinew_0\,|\,\taunewh)}{\vartheta_4(\xinew_0\,|\,\taunew)}\;.
\end{array}
\end{equation}
Note that the weights are chiral, for generic $\xinew_i,\xinew_j$
\begin{equation}
W_\alphanew(\xinew_i,\xinew_j;n_i,n_j)=W_\alphanew(\xinew_j,\xinew_i;n_j,n_i),
\eeq
however
\beq
W_{\alphanew}(\xinew_i,\xinew_j;n_i,n_j)\;\neq\;
W_{\alphanew}(\xinew_i,\xinew_j;n_j,n_i)\;.
\end{equation}
The factor $R$ is given by
\begin{equation}
R\;=\;F_{\alphanew_1}
  F_{\alphanew_3}/F_{\alphanew_1+\alphanew_3}\;,\label{Rnew}
\end{equation}
where all $F_{\alphanew_1}$, $F_{\alphanew_3}$ and
$F_{\alphanew_1+\alphanew_3}$ also implicitly depend on values of
$\xinew_0,\xinew_1,\xinew_2,\xinew_3$, satisfying Eq.\eqref{Q4}.
The corresponding expressions are given by
\begin{equation}
F_{\alphanew_1}^{}\;=\;K_{\alphanew_1}^{}
\frac{P_{\alphanew_1}(\xinew_2,\xinew_3)}{Q_{\alphanew_1}(\xinew_0,\xinew_1)}\;,
\quad
F_{\alphanew_3}\;=\;K_{\alphanew_3}^{}\frac{P_{\alphanew_3}(\xinew_1,\xinew_2)}{Q_{\alphanew_3}(\xinew_0,\xinew_3)}\;,\quad
F_{\alphanew_1+\alphanew_3}\;=\;K_{\alphanew_1+\alphanew_3}^{}
\frac{P_{\alphanew_1+\alphanew_3}(\xinew_0,\xinew_2)}
{Q_{\alphanew_1+\alphanew_3}(\xinew_1,\xinew_3)}\;,\label{Fth}
\end{equation}
where
\begin{equation}
K_\alphanew^{}\;=\;\prod_{n=1}^{N-1}\left[\frac{\vartheta_1'}{2}
\vartheta_1(\frac{\pi}{N}n+\frac{\alphanew}{N})\right]^{n/N}\;.
\end{equation}
Next
\begin{equation}
P_{\alphanew}(\xinew_i,\xinew_j)\;=\;\prod_{n=0}^{N-1}\left[
\frac{\vartheta_1\big(\frac{\pi}{N}(n+\frac{1}{2})+\frac{1}{2N}({\xinew_i-\xinew_j+\alphanew})\big)\:
\vartheta_4\big(\frac{\pi}{N}(n+\frac{1}{2})+\frac{1}{2N}(\xinew_i+\xinew_j+\alphanew)\big)_{\phantom{|}}}
{\vartheta_1\big(\frac{\pi}{N}(n+\frac{1}{2})
+\frac{1}{2N}(\xinew_i-\xinew_j-\alphanew)\big)\,
\vartheta_4\big(\frac{\pi}{N}(n+\frac{1}{2})+\frac{1}{2N}(\xinew_i+\xinew_j-\alphanew)\big)}\right]^{\frac{N-1-2n}{2N}}\;,
\end{equation}
and finally
\begin{equation}
\begin{array}{l}
\ds Q_{\alphanew}\big(\xinew_0,\xinew_k\big)\;=\;
\prod_{n=1}^{N-1}\left[
\vartheta_1\Big(\frac{\pi}{N}n+\frac{1}{2N}(\alphanew-\xinew_0+\xinew_k)\Big)\,
\vartheta_1\Big(\frac{\pi}{N}n+\frac{1}{2N}(\alphanew+\xinew_0-\xinew_k)\Big)
\right.\\
[5mm]
\ds \ \  \ \ \ \ \ \ \ \ \ \ \ \ \ \  \ \ \ \  \ \ \  \ \
\ \ \left.\times\vartheta_4\Big(\frac{\pi}{N}n+
\frac{1}{2N}(\alphanew+\xinew_0+\xinew_k)\Big)\,
\vartheta_4\Big(\frac{\pi}{N}n+\frac{1}{2N}(\alphanew-\xinew_0-\xinew_k)\Big)
\right]^{k/N}\;,
\end{array}
\end{equation}
where $(i,j,k)$ is an arbitrary permutation of $(1,2,3)$.
The above expressions involve theta-functions with
the imaginary period $\tau'/N$, i.e., 
$\vartheta_j(z)=\vartheta_j(z\,|\,\taunewh)$, $j=1,2,3,4$, and
$\vartheta_1'=\vartheta_2\vartheta_3\vartheta_4$ denote 
the corresponding theta-constants. Note that
\beq
P_\alphanew(\xinew_i,\xinew_j)=P_\alphanew(\xinew_j,\xinew_i),\qquad
Q_\alphanew(\xinew_i,\xinew_j)=Q_\alphanew(\xinew_j,\xinew_i)\ .
\eeq

The above solution of the star-triangle equation \eqref{str-sum} contains six 
continuous parameters: the two original spectral parameters
$\alphanew_1,\alphanew_3$, the elliptic period $\tau'$, $\im
\tau'>0$, and the three parameters $\xinew_1,\xinew_2,\xinew_3$. There
is also an integer parameter $N\ge2$, which denotes the number of the discrete
spin states.

\subsection{Kashiwara-Miwa model}
The simplest solution of \eqref{Q4} is
\beq\label{simpsol}
\xinew_j=\pi(\zeta+\nu), \quad \zeta\in{\mathbb Z},\quad
\nu\in\{0,{\textstyle \hf}\},
\qquad j=0,1,2,3,
\eeq
when each factor in \eqref{Q4} equals to one, 
$\Psi_\alphanew(\xinew_i,\xinew_j)\equiv 1$. 
This case leads to the Kashiwara-Miwa (KM) model
\cite{Kashiwara:1986,HY90,Bax02rip}. Since the parameters $\zeta$ and
$\nu$ are the same for all Boltzmann weights, there is no need to
explicitly display the $\phi$-dependence in the star-triangle relation
\eqref{str-sum}. It can be written as  
\begin{equation}\label{str-km}
\begin{array}{l}
\ds \sum_{n_0\in\mathbb{Z}_N} S^{{\KM}}(n_0)\ 
W_{\etanew-\alphanew_1}^\KM(n_1,n_0)\ 
W_{\alphanew_1+\alphanew_3}^\KM(n_2,n_0)\ 
W_{\etanew-\alphanew_3}^\KM(n_3,n_0)\\
[6mm]
\ds \phantom{xxxxxxxxxxxxxxxxx} \;=\;
R^\KM\,
W_{\alphanew_1}^\KM(n_2,n_3)\,
W_{\pi-\alphanew_1-\alphanew_3}^\KM(n_1,n_3)\,
W_{\alphanew_3}^\KM(n_1,n_2)\;.
\end{array}
\end{equation}
where the superscript ``KM'' stands for the Kashiwara-Miwa model. 
Introduce functions 
\begin{equation}
r_{\alphanew}^\KM(n)\;=\;
\prod_{k=1}^n\frac{\vartheta_1(\frac{\pi}{N}(k-\frac{1}{2})-\frac{1}{2N}\alphanew\,|\,\taunewh)}
{\vartheta_1(\frac{\pi}{N}(k-\frac{1}{2})+\frac{1}{2N}\alphanew\,|\,\taunewh)}\,,
\quad t_{\alphanew}^\KM(n)\;=\;
\prod_{k=1}^n\frac{\vartheta_4(\frac{\pi}{N}(k-\frac{1}{2}+\nu)-\frac{1}{2N}\alphanew\,|\,\taunewh)}
{\vartheta_4(\frac{\pi}{N}(k-\frac{1}{2}+\nu)+\frac{1}{2N}\alphanew\,|\,\taunewh)}\,.
\end{equation}
The expressions \eqref{ellipt-cw} give   
\begin{equation}
W_\alphanew^\KM(n_i,n_j)=r_{\alphanew}^\KM(n_i-n_j)
t_{\alphanew}^\KM(n_i+n_j+\zeta)\;,\quad S^\KM(n)=\frac{1}{\sqrt{N}}
\frac{\vartheta_4\big(\frac{\pi}{N}(2n+\zeta+\nu)\,|\,\taunewh)}
{\vartheta_4(\pi\nu\,|\,\taunew\big)}\,,\label{km-weights}
\end{equation}
where have we changed the normalization 
\beq
W_\theta^\KM(0,0)=t^\KM_\theta(\zeta)/t^\KM_\theta(0),\label{km-norm}
\eeq
which is more convenient in this case than \eqref{newnorma}.
The factor $R^\KM$ in \eqref{str-km} 
is obtained by specializing the expression \eqref{Rnew}, 
where $F_\alphanew$ is defined in \eqref{Fth}. Moreover one needs to
take into account the change in normalization \eqref{km-norm} in
comparison with \eqref{newnorma}. In this way one obtains 
\begin{equation}
R^\KM\;=\;F_{\alphanew_1}^\KM
  F_{\alphanew_3}^\KM/F_{\alphanew_1+\alphanew_3}^\KM\;,\label{Rkm}
\end{equation}
where
\begin{equation}\label{fkm}
F_\alphanew^{\KM}\;=\;\prod_{k=1}^{\lfloor\frac{N}{2}\rfloor}
\frac{\vartheta_1\big(\frac{\pi}{N}(k-\frac{1}{2}) +
\frac{1}{2N}\alphanew\,|\,\taunewh\big)}
{\vartheta_1\big(\frac{\pi}{N}k -
\frac{1}{2N}\alphanew\,|\,\taunewh\big)}
\prod_{k=1}^{\lfloor\frac{N-2\nu}{2}\rfloor}
\frac{\vartheta_4\big(\frac{\pi}{N}(k-\hf+\nu)
+\frac{1}{2N}\alphanew\,|\,\taunewh\big)}
{\vartheta_4\big(\frac{\pi}{N}(k+\nu)
-\frac{1}{2N}\alphanew\,|\,\taunewh\big)}\;,
\end{equation}
and the symbol $\lfloor x \rfloor$ denotes the integer part of $x$. 

The resulting solution of the star-triangle relation 
contains three continuous parameters, $\alphanew_1,\alphanew_2$ and 
$\tau'$ and
three discrete parameters $N,\zeta\in {\mathbb Z}$ and $\nu=0,\hf$.
The original Kashiwara-Miwa paper \cite{Kashiwara:1986} 
only deals with the case $\nu=0$.
The case $\nu=\hf$ was considered in \cite{HY90}. The expression
\eqref{fkm} for $\nu=0$ was obtained in \cite{Bax02rip}.

\subsection{Chiral Potts model}

Consider next the limit $\taunew\to\ii\infty$ (trigonometric limit).
In this limit the Jacobi theta-functions 
$\vartheta_3, \vartheta_4$  
of a real argument $x$ 
become identically equal to one,
$\vartheta_3(x|\ii \infty)=\vartheta_4(x|\ii\infty)\equiv1$.
Therefore all functions $\Psi$ in \eqref{newpsi} depend only on the difference
$\xinew_i-\xinew_j$ and thus satisfy the inversion relation:
\begin{equation}
\Psi_\alphanew(\xinew_i,\xinew_j)\;=\; \frac{\cos
\frac{1}{2}(\xinew_i-\xinew_j+\alphanew)}{\cos
\frac{1}{2}(\xinew_i-\xinew_j-\alphanew)}\;=\frac{1}{\Psi_\alphanew(\xinew_j,\xinew_i)}\;.\label{psi-inv}
\end{equation}
Eq.\eqref{Q4}, which determines the value of $\phi_0$,
can now be solved explicitly   
\begin{equation}
\EXP^{\ii \xinew_0}\;=\;\EXP^{\ii(\xinew_1+\xinew_2+\xinew_3)}\;\left(
\frac {\EXP^{-\ii\xinew_1}\sin\alphanew_1 + \EXP^{-\ii\xinew_2}
\sin\alphanew_2 + \EXP^{-\ii\xinew_3}\sin\alphanew_3}
{\EXP^{+\ii\xinew_1} \sin\alphanew_1 + \EXP^{+\ii\xinew_2}
\sin\alphanew_2 + \EXP^{+\ii\xinew_3}\sin\alphanew_3}\right),\label{phi0}
\end{equation}
where $\alphanew_2=\pi-\alphanew_1-\alphanew_3$. 

The star-triangle relation \eqref{str-sum} in this case contains 
only four independent continuous parameters: two differences 
$\xinew_1-\xinew_2$ and $\xinew_1-\xinew_3$ and two spectral parameters
$\alphanew_1$ and $\alphanew_3$. 
We will show that this star-triangle relation  
is equivalent to that of the chiral Potts model 
upon a change of variables.  
The latter has the form \eqref{str-bax} containing three
rapidity variables $p,q$ and $r$. Moreover, the Boltzmann weights
of the chiral Potts model 
depend on one additional parameter --- the modulus of the
rapidity curve (see below). 

Each rapidity variable $p$ is
represented by a three-vector $(x_p, y_p, \mu_p)$ specifying a point
on the algebraic curve, defined by any two of the following equations 
(the third equation follows from the other two)
\begin{equation}
x_\spep^N+y_\spep^N=k(1+x_\spep^Ny_\spep^N)\;,\qquad
k x_p^N=1-k'\mu_p^{-N},\qquad k y_p^N=1-k'\mu_p^N\ .\label{curve}
\end{equation}
where $k$ is a constant (the modulus of the curve) and $k^2+k'^2=1$. 
The rapidities $q$ and $r$ are defined similarly. The Boltzmann
weights of the chiral Pottis model are given by \cite{BPA88} 
\begin{equation}
W_{\spep\speq}^\CP(n)\;=\;\left(\frac{\mu_\spep}{\mu_\speq}\right)^n
\prod_{k=1}^{n}\frac{y_\speq-\omega^kx_\spep}{y_\spep-\omega^kx_\speq}\;,\quad
\overline{W}_{\spep\speq}^\CP(n)=(\mu_\spep\mu_\speq)^n
\prod_{k=1}^n\frac{\omega
  x_\spep-\omega^kx_\speq}{y_\speq-\omega^ky_\spep}\;.
\label{cp-weights}
\end{equation}
where we have assumed the normalization
\beq
W_{pq}(0)=\overline{W}_{pq}(0)=1.\label{cp-norma}
\eeq
These weights satisfy the star-triangle relations
\eqref{str-bax} where the factor $R_{pqr}$ is given by 
\begin{equation}
R^\CP_{pqr}=\frac{f_{qr}^\CP f_{pq}^\CP}{f_{pq}^\CP},\qquad f^\CP_{pq}\;=\;
\prod_{k=1}^{N-1}\left(
\frac{\mu_\speq(1-\omega^k)(t_\spep-\omega^kt_\speq)
(x_\speq-\omega^ky_\spep)}
{\mu_\spep(x_\spep-\omega^kx_\speq)
(y_\spep-\omega^ky_\speq)
(x_\spep-\omega^ky_\speq)}\right)^{k/N}\;.\label{Rpqr-cp}
\end{equation}
Note that for the chiral Potts model the two star-triangle relations in
\eqref{str-bax}
are corollaries of each other. 

Remind, that the variables 
$\alphanew_1,\alphanew_3$ and $\xinew_0,\xinew_1,\xinew_2,\xinew_3$,
appearing in \eqref{str-sum} are constrained by the relation
\eqref{phi0}. Moreover the $\xinew$-variables enter only through
the differences $\xinew_i-\xinew_j$. 
This leaves only four independent continuous degrees of freedom.
Let us parameterize them via three rapidities $p,q,r$ and
the modulus $k$ of the curve \eqref{curve}, 
\beq\begin{array}{c}\ds
\EXP^{\ii\alphanew_1/N}\;=\;\frac{(x_\sper
  y_\sper)^\hf}{(x_\speq y_\speq)^\hf}\;,\quad
\EXP^{\ii\alphanew_3/N}\;=\;\frac{(x_\speq y_\speq)^\hf}{(x_\spep
y_\spep)^\hf}\;,\\[.9cm]
\ds\EXP^{\ii(\xinew_1-\xinew_0)/N}=\frac{(y_q x_r)^\hf}{(x_q
  y_r)^\hf}\,,\qquad
\EXP^{\ii(\xinew_2-\xinew_0)/N}=\omega^\hf \,\frac{(x_p x_r)^\hf}{(y_p
  y_r)^\hf}\,,\qquad
\EXP^{\ii(\xinew_3-\xinew_0)/N}=\frac{(x_p y_q)^\hf}{(y_p
  x_q)^\hf}\,.
\end{array} \label{param}
\eeq
One can easily check that with this parametrization 
the constraint \eqref{phi0} is satisfied by
virtue of the equations defining the algebraic curve \eqref{curve}.
With these substitutions the weights \eqref{ellipt-cw} exactly transform into
the weights of the chiral Potts model, given by \eqref{cp-weights}. 
For instance, the following equalities 
\begin{equation}
W_{\alphanew_1}(\phi_2,\phi_3,n_2,n_3)
\;=\;W_{\speq\sper}^\CP(n_2-n_3)\;,\quad
W_{\pi-\alphanew_1}(\phi_0,\phi_1,n_0,n_1)\;
=\;\overline{W}_{\speq\sper}^\CP(n_0-n_1)\;,
\end{equation}
immediately follow from the relations
\begin{equation}\begin{array}{ll}
\ds
\EXP^{\frac{\ii}{N}(\xinew_2-\xinew_3-\alphanew_1)}
\;=\;\omega^{1/2}\frac{x_\speq}{y_\sper}\;, & \ds
\EXP^{\frac{\ii}{N}(\xinew_2-\xinew_3+\alphanew_1)}
\;=\;\omega^{1/2}\frac{x_\sper}{y_\speq}\;,\\
[5mm]
\ds
\EXP^{\frac{\ii}{N}(\xinew_0-\xinew_1+\alphanew_1)}
\;=\;\frac{y_\sper}{y_\speq}\;, & \ds
\EXP^{\frac{\ii}{N}(\xinew_0-\xinew_1-\alphanew_1)}
\;=\;\frac{x_\speq}{x_\sper}\;.
\end{array}
\end{equation}
which are simple corolaries of \eqref{param}.
Note that the normalization \eqref{newnorma} precicely 
corresponds to \eqref{cp-norma}.
Similarly one obtains, 
\beq
\begin{array}{rclrcl}
\ds W_{\alphanew_1+\alphanew_3}(\phi_2,\phi_0,n_2,n_0)
&=&W_{\spep\sper}^\CP(n_2-n_0)\;,&
\ds W_{\pi-\alphanew_1-\alphanew_3}(\phi_3,\phi_1,n_3,n_1)&
=&\overline{W}_{\spep\sper}^\CP(n_3-n_1)\;,\\[.3cm]
\ds W_{\alphanew_1}(\phi_2,\phi_3,n_2,n_3)
&=&W_{\speq\sper}^\CP(n_2-n_3)\;,& \ds
W_{\pi-\alphanew_1}(\phi_1,\phi_0,n_1,n_0)&
=&\overline{W}_{\speq\sper}^\CP(n_1-n_0)\;.
\end{array}
\eeq
The Boltzmann weight $S(n_0)$ of the central spin in \eqref{str-sum} 
becomes constant, $S(n_0)\equiv1/\sqrt{N}$, while the expressions
\eqref{Fth} transform to  
\begin{equation}
F_{\alphanew_1}\;=\;\frac{1}{\sqrt{N}}f_{\speq\sper}^\CP,\qquad
F_{\alphanew_1+\alphanew_3}\;=\;\frac{1}{\sqrt{N}}f_{\spep\sper}^\CP,\qquad
F_{\alphanew_3}\;=\;\frac{1}{\sqrt{N}}f_{\spep\speq}^\CP,
\end{equation}
so that the factor \eqref{Rnew} exactly reduces to $R_{pqr}^\CP/\sqrt{N}$. 
Thus, there is a precise coincidence with the standard form of
weights and star-triangle relation for the chiral Potts model
\cite{AuY87,Bax02rip}. Note also, that all square roots appearing in
the parametrization \eqref{param}, completely cancel out in the final
expressions for the Boltzmann weights. 

To conclude this Section, 
let us express the rapidities $p,q,r$, and modulus $k$ 
of the curve \eqref{curve} in 
terms of the $\alphanew$'s and $\xinew$'s. It is convenient to define
new variables $\lambda_1,\lambda_2,\lambda_3$, such that 
\beq
\alphanew_1=\lambda_3-\lambda_2,\qquad 
\alphanew_3=\lambda_2-\lambda_1\ . 
\eeq
Eq.\eqref{phi0} can now be re-written as 
\beq
\EXP^{\ii
  \xinew_0}\;=\;\EXP^{\ii(\xinew_1+\xinew_2+\xinew_3)}
\;\frac{U_-}{U_+},\qquad  
U_\pm=\EXP^{\pm\ii\xinew_1}\sin(\lambda_2-\lambda_3) +
  \EXP^{\pm\ii\xinew_2} 
\sin(\lambda_1-\lambda_3)+\EXP^{\pm\ii\xinew_3}\sin(\lambda_1-\lambda_2)\, .
\eeq
Introduce an additional (dependent) variable $\lambda_0$
by a formula obtained from the previous one by interchanging 
all $\xinew$'s and $\lambda$'s,
\beq
\EXP^{\ii
  \lam_0}\;=\;\EXP^{\ii(\lam_1+\lam_2+\lam_3)}
\;\frac{V_-}{V_+},\qquad  
V_\pm=\EXP^{\pm\ii\lam_1}\sin(\xinew_2-\xinew_3) +
  \EXP^{\pm\ii\lam_2} 
\sin(\xinew_1-\xinew_3)+\EXP^{\pm\ii\lam_3}\sin(\xinew_1-\xinew_2)\,.
\eeq
It convenient to define the quantities
\beq
\begin{array}{ll}
\ell_1=(\lam_0+\lam_1-\lam_2-\lam_3)/2N\,, \qquad 
&f_1=(\xinew_0+\xinew_1-\xinew_2-\xinew_3)/2N\,,\\[.3cm] 
\ell_2=(\lam_0+\lam_2-\lam_1-\lam_3)/2N\,, \qquad 
&f_2=(\xinew_0+\xinew_2-\xinew_1-\xinew_3)/2N\,,\\[.3cm] 
\ell_3=(\lam_0+\lam_3-\lam_1-\lam_2)/2N\,, \qquad 
&f_3=(\xinew_0+\xinew_3-\xinew_1-\xinew_2)/2N\,.
\end{array}
\eeq
Then, inverting \eqref{param}, one obtains
\beq
\begin{array}{rclrclrcl}
x_p&=&\EXP^{\ii(\ell_1-f_1)},\qquad 
&x_q&=&\EXP^{\ii(\ell_2+f_2)},\qquad 
&x_r&=&\EXP^{\ii(\ell_3-f_3)},\\[.3cm]
y_p&=&\omega^\hf\,\EXP^{\ii(\ell_1+f_1)},\qquad 
&y_q&=&\omega^\hf\,\EXP^{\ii(\ell_2-f_2)},\qquad 
&y_r&=&\omega^\hf\,\EXP^{\ii(\ell_3+f_3)},
\end{array}
\eeq
and
\beq
k^2=({V_+ V_-})/({U_+U_-})\ .
\eeq
For real $\xinew$'s and $\lam$'s the quantities $V_+$ and $V_-$ are
complex conjugate to each other (the same  is true for $U_+$ and $U_-$).
In this case the last formula can be written as  
\begin{equation}
k\;=\;\left|\frac{\EXP^{\ii\lambda_{1}}
\sin(\xinew_2-\xinew_2) + \EXP^{\ii\lambda_{2}}
\sin(\xinew_1-\xinew_3) + \EXP^{\ii\lambda_{3}}
\sin(\xinew_1-\xinew_2)} {\EXP^{\ii\xinew_1}
\sin(\lambda_2-\lambda_3) + \EXP^{\ii\xinew_2}
\sin(\lambda_1-\lambda_3) + \EXP^{\ii\xinew_3}
\sin(\lambda_1-\lambda_2)}\right|\;.
\end{equation}
It is worth noting an explicit, but rather unexpected,  
symmetry of the above expressions upon interchanging $\xinew$'s and $\lam$'s.
It would be interesting to understand reasons behind this phenomenon. 

The results of this Section are closely related to the description
of the chiral Potts model previously obtained by one of us in
\cite{Baz08}.

\section{Conclusion}

We have obtained a new solution \eqref{w-main} of the star-triangle relation 
\eqref{str-main}, which is expressed in terms of the elliptic
$\Gamma$-function. The solution defines a perfectly physical
two-dimensional solvable model of statistical mechanics 
with positive
Boltzmann weights. Its partition function is defined in
\eqref{Z-def2}. This is an Ising-type model with continuous real spins
$0\le x_i<2\pi$, which can be regarded as angle variables on a circle. 
The model contains two temperature-like variables $\p$ and $\q$; 
see \eqref{reg12} for a definition of the main physical regimes.

We would like to stress that 
the formula \eqref{w-main} provides a {\em master solution}
of the star triangle-relation in the sense that 
it contains as special cases all continuous and discrete spin
solution of this relation that were previously
known\footnote{See footnote 1 on page 2.}. 
In particular, in Sections~4.2 and 4.3 
we explicitly demonstrate how to obtain the Kashiwara-Miwa and
chiral Potts models. 
We show that these models are particular cases of
a more general ``hybrid'' model, which couples a classical integrable
system involving continuous lattice spins 
to an Ising-type model of statistical
mechanics with discrete spins, taking $N\ge2$ different values.   
This hybrid model arises from the general case when one of
the temperature-like parameters approaches a root of unity, $\q^N=1$. 
The required limiting procedure is considered in Section~3.
To formulate this hybrid model on an arbitrary planar graph (of the type
discussed in Section~2.1) one needs to assign two types
of spin variables to lattice sites: 
classical variables $\phi_i$, satisfying equations
\eqref{vareq}, and discrete spin variables $n_i\in{\mathbb Z}_N$.
Then
one needs to solve the equations \eqref{vareq} with fixed boundary
conditions and determine the 
variables $\phi_i$ for all internal sites of the lattice.
After that the discrete spin model is defined by the weights
$W_\theta(\phi_i,\phi_j,n_i,n_j)$ and $S(\phi_0,n_0)$ given by
\eqref{ellipt-cw}. In general, it is a spatially-inhomogeneous model
where the Boltzmann weight vary across the lattice,  since they depend
on the variables $\xinew_i$. Nevertheless, it is an integrable model.
In particular, its partition function also possesses 
the Baxter's $Z$-invariance property of Section~2.1 
by virtue of the star-triangle relation \eqref{str-sum} and the
variational equations \eqref{vareq} for the classical variables
$\xinew_i$. 

It appears that our master solution \eqref{w-main} is also deeply connected 
with many other important models of statistical
mechanics, whose consideration goes beyond the scope of this paper. 
We hope to unravel these connections in the future.

\section*{Acknowledgements} 

The authors thank L.D.~Faddeev for interesting comments and
R.J.~Baxter for reading the manuscript and numerous valuable remarks
and suggestions, which were taken into account in the final version of
the paper. We are also indebted to H.~Au-Yang, M.T.~Batchelor,
G.P.~Korchemsky, 
V.V.~Mangazeev and J.H.H.~Perk for their interest to this work and
useful discussions.

\app{Energy functional in original variables}
\label{appA}
In the original variables $\xi_j,\alpha,\tau$ the asymptotic of
\eqref{w-weight} and \eqref{s-weight} is
 \bea
\log {\cal W}_\alpha(x_1,x_2)&=&-\frac{1}{\varepsilon}\,\widetilde{\cal
    L}(\alpha\,|\,\xi_1,\xi_2) +O(1)\ ,\label{w-ass-old}\\
    [.3cm]
\log \s(x_i)&=&-\frac{1}{\varepsilon}\,\widetilde{\cal C}(\xi_i)
-\frac{1}{2}\log{\varepsilon}+O(1)\label{s-ass-old}
 \eea
where $\widetilde{\cal L}(\alpha\,|\,\xi_1,\xi_2)\equiv {\cal L}(\alphanew\,|\,\xinew_1,\xinew_2)$
and $\widetilde{\cal C}(\xi_i)\equiv {\cal C}(\xinew_i)$, $\xi,\alpha$ are related with $\xinew,\alphanew$ by (\ref{newvars}) and
${\cal L}$ and ${\cal C}$ have appeared in (\ref{w-ass},\ref{s-ass}). Expressions for $\widetilde{\cal L}$ and $\widetilde{\cal S}$ are then given by
\begin{equation}
\widetilde{\cal L}(\alpha\,|\,\xi_1,\xi_2)=\ii N \left\{
\mathop{\int}_{\!\!\!\!\!0}^{\;\;\;\;\xi_1-\xi_2} dz
\log\frac{\vartheta_3\big(\frac{N}{2}(z-\ii\alpha)\,|\,N\tau\big)}{\vartheta_3\big(\frac{N}{2}(z+\ii\alpha)\,|\,N\tau\big)}
+\mathop{\int}_{\!\!\!\!\pi/N}^{\;\;\;\;\xi_1+\xi_2} dz
\log\frac{\vartheta_3\big(\frac{N}{2}(z-\ii\alpha)\,|\,N\tau\big)}{\vartheta_3\big(\frac{N}{2}(z+\ii\alpha)\,|\,N\tau\big)}\right\}\;,
\end{equation}
and therefore
\begin{equation}
\begin{array}{l}
\ds\widetilde{\cal
L}(\eta-\alpha\,|\,\xi_1,\xi_2)=\frac{\pi^2}{2}-(N\xi_1)^2-(N\xi_2)^2\\
[5mm]
\ds + \ii N \left\{
\mathop{\int}_{\!\!\!\!\!0}^{\;\;\;\;\xi_1-\xi_2} dz
\log\frac{\vartheta_1\big(\frac{N}{2}(\ii\alpha+z)\,|\,N\tau\big)}{\vartheta_1\big(\frac{N}{2}(\ii\alpha-z)\,|\,N\tau\big)}
+\mathop{\int}_{\!\!\!\!\pi/N}^{\;\;\;\;\xi_1+\xi_2} dz
\log\frac{\vartheta_1\big(\frac{N}{2}(\ii\alpha+z)\,|\,N\tau\big)}{\vartheta_1\big(\frac{N}{2}(\ii\alpha-z)\,|\,N\tau\big)}\right\}\;.
\end{array}
\end{equation}
Finally, the function ${\cal C}(\xi_i)$ has the period
$\ds\frac{\pi}{N}$ and on the interval $\ds 0<\xi_i<\frac{\pi}{N}$
it is given by
\begin{equation}
\widetilde{\cal C}(\xi_i)\;=\;2 \Big(N\xi_i-\frac{\pi}{2}\Big)^2\;.
\end{equation}
Note that the
quadratic in $\xi$ terms cancel out from the energy functional and,
therefore, do not contribute to the variational equations \eqref{q4-eq0} for
square lattice.  
For $\rho=1$ and $T=\varepsilon$, the energy ${\cal E}$ 
in (\ref{z-low}) is given by
\begin{equation}
{\cal E}(X)\;=\;\sum_{(ij)}\widetilde{\cal L}(\alpha\,|\,\xi_i,\xi_j) 
+ \sum_{(kl)}\widetilde{\cal L}(\eta-\alpha\,|\,\xi_k,\xi_l)
+\sum_{m} \widetilde{\cal C}(\xi_m)\;,
\end{equation}
where the first sum is taken over all horizontal edges $(ij)$, the
second over all vertical edges $(kl)$ and the third over all internal
sites $m$ of the lattice.  
(\ref{z-main}).

%\bibliography{total33,elliptic}
%\bibliographystyle{vvb-bibstyle}

\newcommand\oneletter[1]{#1}

\end{document}